\documentstyle[aps,preprint]{revtex}
\begin{document}
\draft
\tightenlines
 
\title{How to simulate the quasi-stationary state}
\author{Marcelo Martins de Oliveira$^*$ and Ronald Dickman$^\dagger$ 
}
\address{
Departamento de F\'{\i}sica, ICEx,
Universidade Federal de Minas Gerais,\\
30123-970
Belo Horizonte - Minas Gerais, Brasil\\
}

\date{\today}

\maketitle
\begin{abstract}
For a large class of processes with an absorbing state, 
statistical properties of the surviving sample
attain time-independent values in the
{\it quasi-stationary} (QS) regime.
We propose a practical simulation method for studying quasi-stationary
properties, based on the equation of motion governing the QS distribution.
The method is tested in applications to the contact process.
At the critical point, our method is about an order of magnitude more
efficient than conventional simulation.
\end{abstract}


\noindent {\small $^*$electronic address: mancebo@fisica.ufmg.br}

\noindent {\small $^\dagger$electronic address: dickman@fisica.ufmg.br}

\newpage
%\section{Introduction}

Stochastic processes with an absorbing state arise frequently in
statistical physics \cite{vankampen,gardiner}, epidemiology
\cite{bartlett}, and related fields.
In autocatalytic processes (in lasers, chemical reactions, 
or surface catalysis, for example), and population models,
a population
of $n \geq 0$ individuals goes extinct when the
absorbing state, $n\!=\!0$, is reached.  
Phase transitions to an absorbing state in spatially extended
systems, exemplified by the contact process \cite{harris,liggett}, 
are currently of great interest
in connection with self-organized criticality \cite{socbjp}, 
the transition to turbulence \cite{bohr}, and 
issues of universality in nonequilibrium
critical phenomena \cite{marro,hinrichsen,odor04}.  Such models are 
frequently studied using deterministic mean-field equations, Monte Carlo 
simulation, and renormalization group analyses.  
 
It is desirable to develop additional methods for studying
processes with absorbing states.  In this work we study models that 
admit an active (nonabsorbing) stationary 
state in the infinite-size limit, but which always, for a
finite system size, end up in the absorbing state.
The {\it quasi-stationary} (QS) distribution for such a system  
provides a wealth of information about its
behavior.   (In fact, simulations of ``stationary" properties of lattice 
models with an absorbing state actually study the quasi-stationary regime,
given that the only true stationary state for a finite system is the absorbing 
one.)  

In particular, it would be valuable to have a simulation 
method that yields quasi-stationary properties directly.
Currently available methods involve a somewhat complicated procedure
for determining QS properties: a large sample of independent realizations
are performed, and the mean $\phi(t)$ 
of some property (for example the order parameter)
is evaluated over the surviving realizations at time $t$. 
At short times times $\phi(t)$
exhibits a transient as it relaxes toward the QS regime; at long
times it fluctuates wildly as the surviving sample decays.  Normally one
is able to identify an intermediate regime free of transients and
with limited fluctuations, which can be used to estimate the QS value
of $\phi$.  This method requires careful scrutiny of the data and is not
always free of ambiguity \cite{heger}.

A number of strategies have been devised to circumvent the
difficulties associated with simulating models exhibiting absorbing states.
One involves replacing the absorbing state with a {\it reflecting} 
boundary in configuration space \cite{mnrst}.  In another 
approach the activity is fixed
at a nonzero value, in a ``constant coverage" simulation \cite{brosilow}
or a ``conserved" version of the model \cite{tomeoliveira}.   If one includes a weak 
external source of activity, $h$, the zero-activity state is no longer absorbing,
but it is possible to obtain information on critical behavior by 
analyzing scaling properties as $h \to 0$ \cite{heger}.
A further possibility is to clone surviving realizations of the process, 
enriching
the sample to compensate for attrition as the survival probability 
decays \cite{winners}.  While all of these approaches are useful, none 
affords direct, unbiased sampling of the QS state of the original
process.

For models without spatial structure, such as uniformly distributed 
populations or well stirred chemical reactors,
the full QS distribution can be found from the master equation,
via recurrence relations or an iterative 
scheme \cite{RDVIDIGAL,intme}.
In models with spatial 
structure, typified by nearest-neighbor interactions on a lattice, 
mean-field like approximations to the QS distribution have been
derived, but descriptions
in terms of one or a few random variables cannot capture critical 
fluctuations.  The simulation method developed here does not
suffer from this limitation.  It provides a sampling of the QS 
probability distribution
just as conventional Monte Carlo simulation 
does for the equilibrium distribution.  

We begin reviewing the definition  
of the quasi-stationary distribution.  
Consider a continuous-time Markov process $X_t$ taking values 
$n = 0, 1, 2,...,S$, 
with the state $n\!=\!0$ absorbing.  
We use $p_n(t)$ to denote the probability that $X_t = n$, given some initial
state $X_0$. The survival probability
 $P_s(t) = \sum_{n \geq 1} p_n(t)$
is the probability that the process has not become trapped
in the absorbing state up to time $t$.
We suppose that as $t \to \infty$ the $p_n$, normalized by
the survival probability $P_s(t)$, attain a time-independent 
form.
The quasi-stationary distribution $\overline{p}_n$ is then 
defined via

\begin{equation}
\overline{p}_n  = \lim_{t \to \infty} \frac{p_n (t) }{P_s (t)} 
,\;\;\;\; (n \geq 1),
\label{qshyp}
\end{equation}
with $\overline{p}_0 \equiv 0$. The QS distribution is
normalized so:
\begin{equation}
\sum_{n \geq 1} \overline{p}_n = 1.
\label{norm}
\end{equation}

We now discuss the theoretical basis for our simulation method. 
The QS distribution is the 
stationary solution to
the following equation of motion \cite{RDVIDIGAL} (for $n > 0$)

\begin{equation}
\frac{d q_n}{dt} = -w_n q_n + r_n + r_0 q_n \;,
\label{qme}
\end{equation}
where $w_n = \sum_m w_{m,n}$ is the total rate of transitions out of state
$n$, and $r_n = \sum_m w_{n,m} q_m$ is the flux of probability into this
state.  To see this, consider the master equation (Eq. (\ref{qme})
without the final term) in the QS regime.  Substituting
$q_n(t) = P_s(t) \overline{p}_n$, and noting that in the QS regime
$d P_s/dt = -\overline{r}_0 = - P_s \sum_m w_{0,m} \overline{p}_m$, 
we see that the r.h.s. of Eq. (\ref{qme}) is identically zero
if $q_n = \overline{p}_n$ for $n \geq 1$.
The final term in Eq. (\ref{qme}) represents a redistribution of the 
probability $r_0$ (transfered to the
absorbing state in the original master equation), 
to the nonabsorbing subspace.
Each nonabsorbing state receives a share
equal to its probability.  

Although Eq. (\ref{qme})
is not a master equation (it is nonlinear in the probabilities $q_n$),
it does suggest a simulation scheme for sampling the QS distribution.
In a Monte Carlo simulation
one generates a set of realizations of a stochastic 
process.  In what follows we call a simulation of the original 
process $X_t$ (possessing an absorbing state) a {\it conventional}
simulation.
Our goal is to define a related process, $X_t^*$, whose {\it stationary} 
probability distribution is the {\it quasi-stationary} distribution 
of $X_t$.
(Note that in order to have a nontrivial stationary
distribution, $X_t^*$ cannot
possess an absorbing state.)  
The probability distribution of $X_t^*$ is governed by Eq. (\ref{qme}),
which implies that for $n>0$ (i.e., away from the absorbing
state), the evolution of $X_t^*$ is identical to that of $X_t$.
(Since  Eq. (\ref{qme}) holds for $n > 0$, the process
$X_t^*$ must begin in a non-absorbing state.)
When $X_t$ enters the absorbing state, however, $X_t^*$ instead
jumps to a nonabsorbing one, and then resumes its ``usual" evolution
(with the same transition probabilities as $X_t$), until such time
as another visit to the absorbing state is imminent.

A subtle aspect of Eq. (\ref{qme}) is that
the distribution $q_n$ is used to determine the value of
$X_t^*$ when $X_t$ visits the absorbing state.  Although one
has no prior knowledge of $q_n$ (or its long-time limit,
the QS distribution $\overline{p}_n$),
one can, in a simulation, use the 
history $X_s^*$ ($0 < s \leq t$) up to time $t$, to
{\it estimate} the $q_n$.  
This is done by saving 
(and periodically updating) a sample $n_1, n_2, ..., n_M$ of the states 
visited.
(If the state space is characterized by a small set of variables, 
one may instead accumulate a histogram $H(n)$ giving the time spent 
in state $n$.) 
As the evolution progresses, 
$X_s^*$ will visit states according
to the QS distribution.  
We therefore update the sample $\{n_1,n_2,...,n_M\}$ 
by periodically replacing one of these configurations 
with the current one.  
In this way the distribution for the process $X_t^*$ 
(and the sample drawn from it),
will converge to the QS distribution (i.e., the stationary
solution of Eq. (\ref{qme})) at long times.
Summarizing, the simulation process $X_t^*$ has the same dynamics as 
$X_t$, except that when a transition to the absorbing 
state is imminent, $X_t^*$ is placed in a nonabsorbing state, 
selected at random from a sample over the history of the realization.
In effect, the nonlinear term in Eq. (\ref{qme}) is represented as a 
{\it memory} in the simulation.

To explain how our method works in practice, 
we detail its application to the
{\it contact process} (CP) \cite{harris,liggett,marro}.
In the CP, 
each site $i$ of a lattice is 
either occupied ($\sigma_i (t)= 1$),
or vacant ($\sigma_i (t)= 0$).  Transitions from $\sigma_i = 1$ to 
$\sigma_i = 0$ occur at a rate of unity, independent of the neighboring sites.  
The reverse transition can only occur if at least one neighbor is 
occupied: the transition from $\sigma_i = 0$ to $\sigma_i = 1$ 
occurs at rate 
$\lambda r$, where $r$ is the fraction of nearest neighbors of site $i$ 
that are occupied; thus the state $\sigma_i = 0$ for all $i$ is absorbing.
($\lambda $ is a control parameter governing the rate of spread of
activity.)
The order parameter $\rho$ is the fraction of occupied sites.

To begin we study the  
contact process on a {\it complete graph} (CPCG),  
in which the rate for a vacant site to turn occupied is $\lambda$ times the 
fraction of {\it all} sites that are occupied, rather than the fraction of 
nearest neighbors.  Since each site interacts equally with all others, 
all pairs of sites are neighbors,
defining a complete graph.
(The critical value in this case is $\lambda_c=1$; the stationary
value of $\rho$ is zero for $\lambda < \lambda_c$.)  
The state of the process is specified by a single variable $n$: 
the number of occupied sites.  This is a one-step process with nonzero 
transition rates
\begin{equation}
W_{n-1,n} = n
\label{cpcg1}
\end{equation}
\begin{equation}
W_{n+1,n} = \lambda \frac{n}{L} (L-n)
\label{cpcg2}
\end{equation}
on a graph of $L$ sites.  In Ref. \cite{RDVIDIGAL} the exact
QS distribution for the CPCG is obtained via a set of recurrence
relations.

We simulate the quasi-stationary state of the CPCG by realizing the process
corresponding to the transition rates of Eqs. (\ref{cpcg1}) 
and (\ref{cpcg2}) and maintaining a
list of $M = 10^4$ states.  Each time a (nonabsorbing) state is visited, 
we update the list with probability $\gamma \Delta t$
where $\Delta t = 1/w_n$ is the mean duration of this state. 
Whenever a transition to the absorbing state ($n=0$) is generated,
we instead select a state from the list.
The results for a system of 100 sites, 
using $\gamma = 0.5$, are
shown in Fig. 1, illustrating excellent agreement
with the exact QS distribution obtained 
in Ref. \cite{RDVIDIGAL}.
The simulation result is in good agreement with the exact result even
for $\lambda = 0.1$, deep in the subcritical regime, 
in which the lifetime of the original process is very short.

Next we turn to the one-dimensional contact process (i.e, the model defined 
above on a ring of $L$ sites).  Although no exact results 
are available, the model has been studied intensively via series
expansion and Monte Carlo simulation.
In the QS simulations we use a list size $M= $ 
2 $\times 10^3$ - $10^4$, depending on the lattice size.  
The process is simulated in runs of 10$^8$ or more time steps.
As is usual, annihilation events are chosen with probability $1/(1+\lambda)$ and
creation with probability $\lambda/(1+\lambda)$.  A site is chosen from
a list of currently occupied sites, and, in the case of annihilation, is 
vacated, while, for creation events, a nearest-neighbor site is selected 
at random and, if it is currently vacant, it becomes occupied.  The time 
increment associated with each event is $\Delta t = 1/N_{occ}$, where 
$N_{occ}$ is the number of occupied sites just prior
to the attempted transition \cite{marro}.

In the initial phase of the evolution, the list of saved configurations 
is augmented whenever the time $t$ increases by one,  
until a
list of $M$ configurations is accumulated.  
From then on, 
we update the list 
(replacing a randomly selected entry with the current
configuration),
with a certain probability $p_{rep}$,  
whenever $t$ advances by one unit. 
A given configuration therefore remains on the list for a mean time 
of $M/p_{rep}$.  (Values of $p_{rep}$ in the range $10^{-3} - 10^{-2}$ 
are used; the results appear to be insensitive to the precise choice.)

Fig. 2 shows that our method reproduces the order parameter 
$\rho$ obtained via conventional simulations. 
In Fig. 3 the QS and conventional results for the moment ratio 
$m = \langle \rho^2 \rangle/\langle \rho \rangle^2$ are compared.
As discussed in Ref. \cite{RDVIDIGAL},
the {\it lifetime} of the QS state is given by $\tau = 1/\overline{p}_1$.  
The lifetime obtained in QS simulations compares well
with the lifetime obtained via conventional simulation
(using $P_s \sim \exp(-t/\tau)$), as shown
in Fig. 4.   Detailed comparison shows that there is no significant
difference between the QS and conventional simulation results.
For $\lambda < 3$, conventional simulations are  
subject to relatively large uncertainties, since almost all
realizations become trapped in the absorbing state before the 
quasi-stationary regime is attained.
At the critical point, quasi-stationary simulations require
about an order of magnitude less cpu time than conventional
simulations, to achieve results of the same precision.
(Below $\lambda_c$ the savings is even greater; well above
$\lambda_c$ the two methods are equally efficient, since visits to the 
absorbing state are extremely rare.)

As a further application of our method we
study the critical CP ($\lambda_c = 3.297848$) on rings of
$80, 160,...,1280$ sites.
We perform between 5 and 30 realizations, each extending to
$5 \times 10^8$ time steps.
Results from the first 10$^7$ time steps
($2 \times 10^7$ for $L=1280$) are discarded from the averages; 
convergence to the QS 
state occurs on a considerably shorter time scale.
Varying the list size $M$ and the
replacement probability $p_{rep}$, we obtain results consistent with
conventional simulations for $M = 1000$ or greater, and $p_{rep} \leq
10^{-2}$.  Smaller list sizes and/or larger values of $p_{rep}$ yield 
results slightly, but significantly, different from previous studies.
On the other hand, there seems to be little point in using a list
size much greater than the number of reinitializations.  
In practice the best approach is to use $M = 10^3$ or more, and to
perform simulations with several values of $p_{rep}$,
to verify that the results show no significant dependence on this parameter.

Our results permit us to refine the estimate for the 
moment ratio $m = \langle \rho^2 \rangle/\langle \rho \rangle ^2$ ($\rho$ 
denotes the fraction of occupied sites).
This quantity is analogous to Binder's reduced fourth cumulant 
\cite{binder} at an equilibrium critical point: the curves 
$m(\lambda,L)$ for various $L$ cross near $\lambda_c$ (the 
crossings approach $\lambda_c$ as $L$ increases), 
so that $m$ assumes a universal
value $m_c$ at the critical point.  
In Ref. \cite{rdjaff} the estimate $m_c = 1.1736(2)$ for the critical 
one-dimensional CP was obtained
from simulations of systems of up to 320 sites.
Using the present method we reproduce and
refine the results of \cite{rdjaff}. 
Based on our data for $L \leq 640$ we estimate $m_c = 1.17370(2)$.

At the critical point, the order parameter 
is expected to
decay as a power law, $ \rho \sim L^{-\beta\/\nu_\perp}$.  
The QS simulation data (for $L=80$ - 1280)
follow a power law with $\beta\/\nu_\perp = 0.2525(5)$, 
compared with the literature value of 0.2521.  
The lifetime $\tau \sim L^{\nu_{||}/\nu_\perp}$ at the critical point; 
our simulation results yield $\nu_{||}/\nu_\perp = 1.576(2)$ while
the standard value is 1.5807.  We find that including a correction to scaling
scaling term of the form
$\tau \sim L^{\nu_{||}/\nu_\perp}(1 - bL^{-\phi})$ yields a substantially
better fit (the sum of the squares of the errors is reduced by
a factor of 2) if we use $\phi = 0.2$.  The best-fit parameters are
$b=0.069$ and $\nu_{||}/\nu_\perp =1.581(2)$.  (Precise
determination of this correction to scaling term  
will require high precision data for a larger set of system sizes,
a task we defer to future work.)  Summarizing, the QS simulation
method yields results fully consistent with conventional simulation, and
with established scaling properties, when applied to the contact process
on a ring.

It is interesting to compare the QS distribution with that obtained
using a reflecting boundary at $n=0$, as was used in \cite{mnrst}.
In the case of the CP on a complete graph the stationary distribution
with a reflecting boundary (RB) is given in \cite{RDVIDIGAL}, where it is
called the ``pseudo-stationary" distribution.  We find that
for large systems, in the 
active phase ($\lambda$ well above $\lambda_c$) the RB and QS
distributions are essentially the same, but that nearer (and below)
the transition the RB distribution yields a much higher probability
for states near $n=1$ than does the QS.  
The reflecting boundary is equivalent to a modified process $X_t^*$
which, when a visit to the absorbing state is 
imminent, is always reset to the previous configuration. 
Since the latter is but one step removed from the absorbing state,
the buildup of probability in the vicinity is not surprising.
In general, we should expect the QS and RB distributions to be in good
accord when the lifetime of the process is reasonably long, since this
implies a small QS probability in the vicinity of the absorbing state.

In summary, we have devised and tested a simulation method for
quasi-stationary properties of models with an absorbing state.
The method is easy to implement, and
yields reliable results in a fraction of the time required 
for conventional simulations in the critical regime, of prime
interest in the context of scaling and universality.
It also opens the possibility of investigating QS 
properties in the subcritical regime, which is essentially inaccessible
to conventional simulations.
We expect the method to be applicable to many problems currently
under investigation, such as branching-annihilating random walks, 
conserved sandpiles, and stochastic population models.

\vspace{1em}

\noindent{\bf Acknowledgment}

This work was supported by CNPq and FAPEMIG, Brazil.

\newpage

FIGURE CAPTIONS
\vspace{1em}

\noindent FIG. 1. QS distributions via recurrence relations
(solid lines) and QS simulation (symbols) for the CP on a complete
graph of 100 sites. $\lambda = 0.5$, 1.0 and 1.5 (left to right).
%cpcg.eps
\vspace{1em}

\noindent FIG. 2.Quasi-stationary density $\rho$ in the 
one-dimensional CP. Open symbols: QS simulation, $L=20$; filled
symbols: QS simulation, $L=200$.  Solid lines represent
results of conventional simulations.
%comprho.eps
\vspace{1em}

\noindent FIG. 3. Quasi-stationary moment ratio $M$ in the 
one-dimensional CP.  Symbols as in Fig. 2.
%compm.eps
\vspace{1em}

\noindent FIG. 4. Quasi-stationary lifetime $\tau$ in the 
one-dimensional CP.  Symbols as in Fig. 2.
%comptau.eps
\vspace{1em}

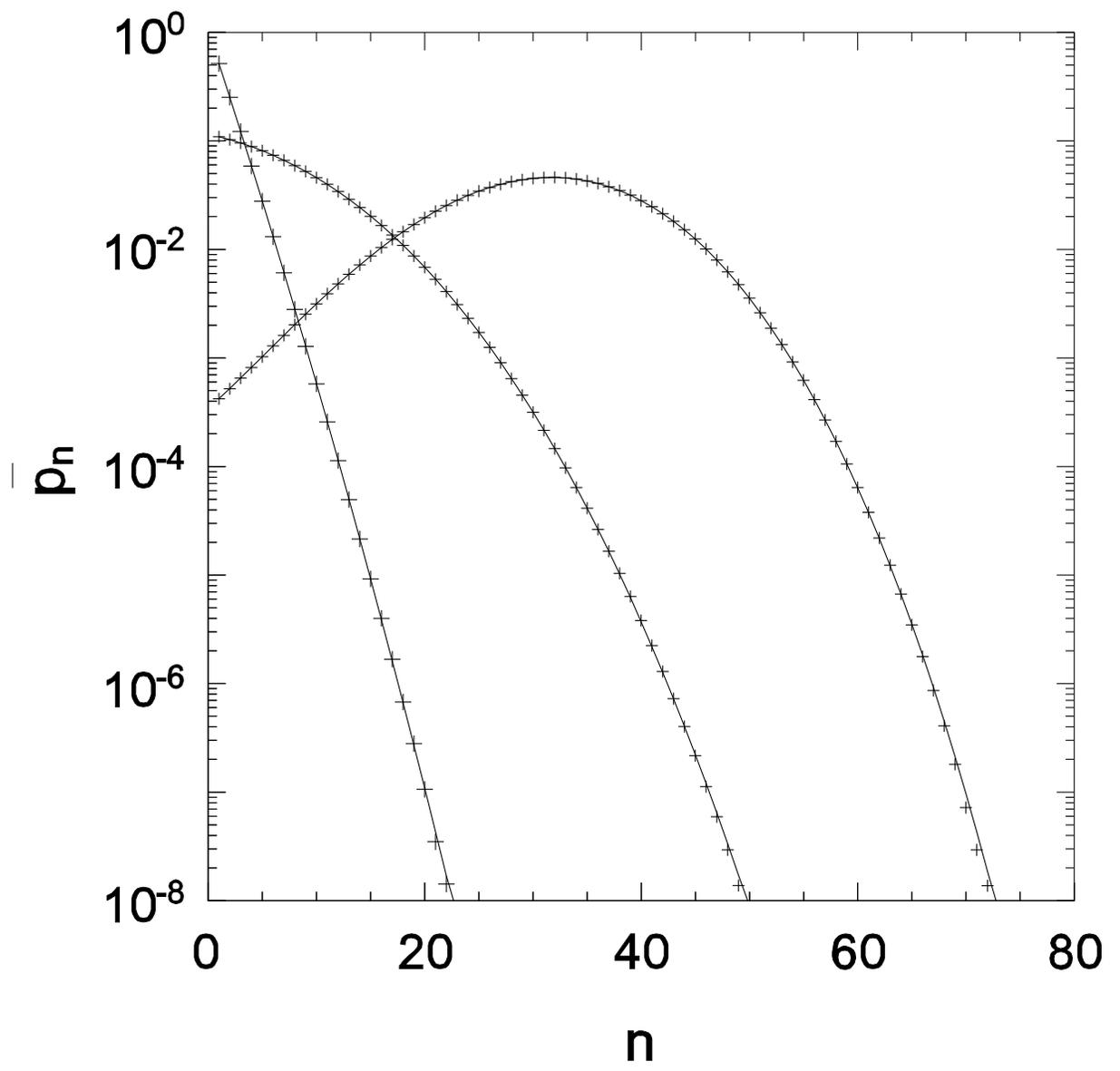
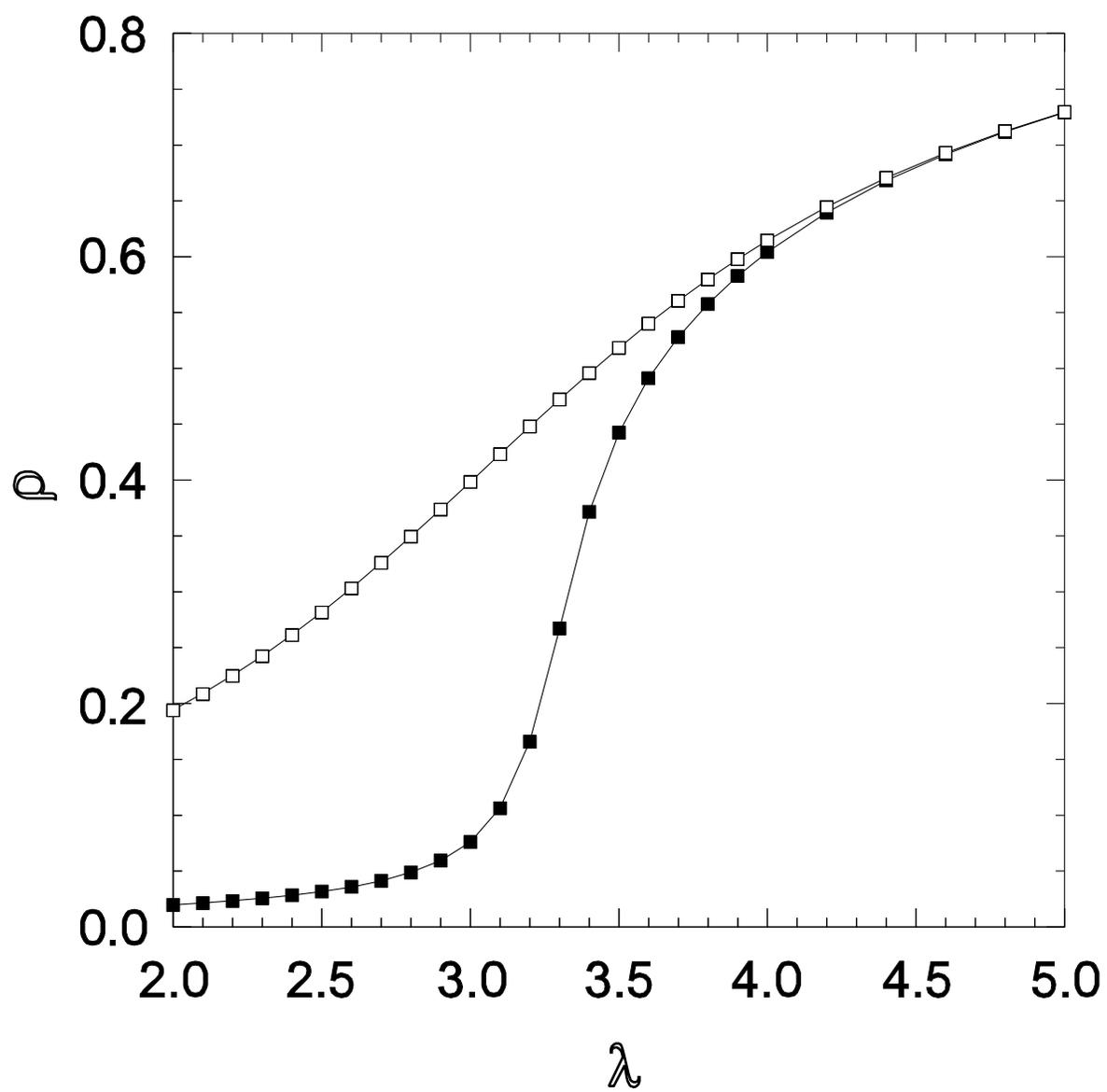
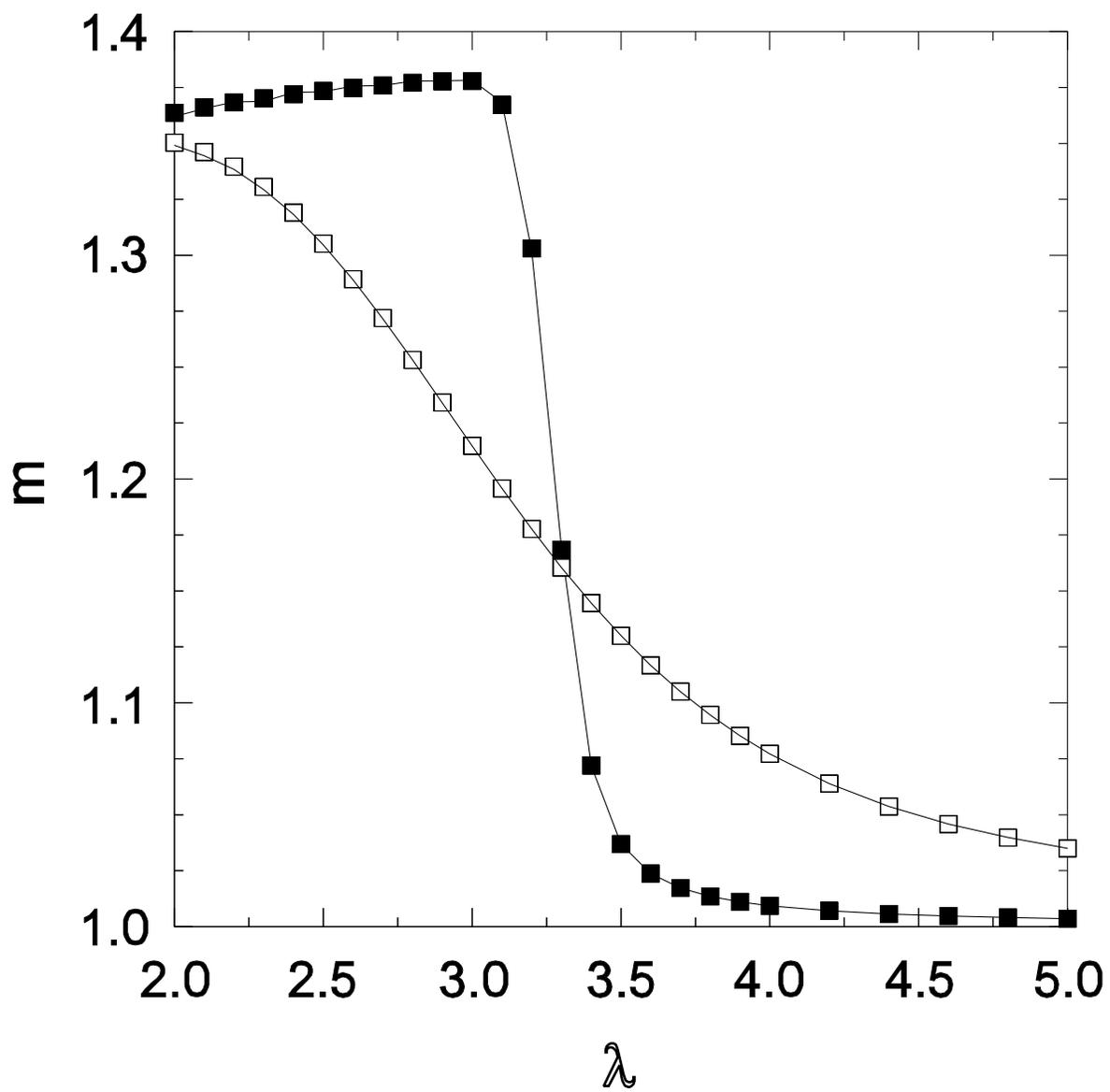
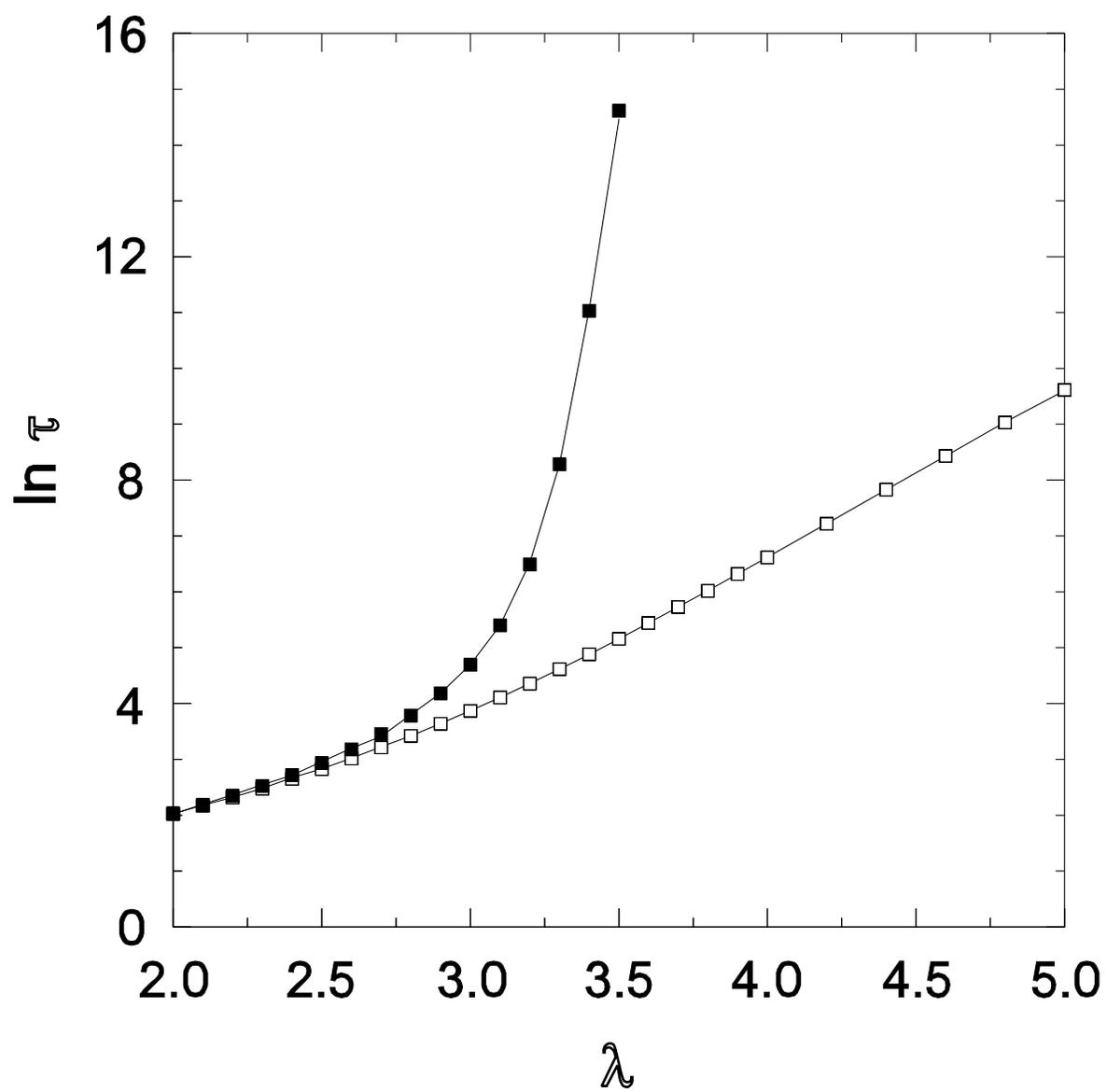
\end{document}